\documentclass[aps,pra,reprint,linenumber,superscriptaddress,nofootinbib]{revtex4-1}
\usepackage{amssymb}
\usepackage{amsmath}
\usepackage{graphicx}
\usepackage{xcolor}
\usepackage{bm}
\usepackage{epstopdf}
\usepackage{float}
\usepackage[colorlinks=true,linkcolor=red,citecolor=blue,urlcolor=red]{hyperref}

\begin{document}

\title{Effect of resonance magnetic perturbation on edge-core turbulence spreading in a tokamak plasma}

\author{Guangzhi Ren}
\affiliation{Key Laboratory of Materials Modification by Laser, Ion, and Electron Beams (Ministry of Education), School of Physics, Dalian University of Technology, Dalian 116024, China}

\author{Lai Wei}
\email{laiwei@dlut.edu.cn}
\affiliation{Key Laboratory of Materials Modification by Laser, Ion, and Electron Beams (Ministry of Education), School of Physics, Dalian University of Technology, Dalian 116024, China}

\author{Zheng-Xiong Wang}
\email{zxwang@dlut.edu.cn}
\affiliation{Key Laboratory of Materials Modification by Laser, Ion, and Electron Beams (Ministry of Education), School of Physics, Dalian University of Technology, Dalian 116024, China}

\author{Jiquan Li}
\affiliation{Southwestern Institute of Physics, Sichuan, Chengdu 610041, China}

\date{\today}

\begin{abstract}

Turbulence spreading from edge to core region with resonance magnetic perturbation (RMP) is investigated using an electromagnetic Landau-fluid model in toroidal geometry.
When RMP field with appropriate amplitude are employed in the simulation, long wavelength fluctuations around the resonance surface are excited due to the forced magnetic reconnection.
Strong shear flow at the magnetic island separatrix are observed, and break the radial elongated vortex structures of the turbulent fluctuation.
The inward flux could be blocked by this shear flow and the saturation level in the core region declines.

\end{abstract}

\maketitle

\section{Introduction}

Turbulence transport \cite{horton_1999,doyle_2007} in the toroidal plasmas, mainly induced by the drift wave type instabilities, is one of the key topics in magnetic fusion research. 
As one typical transport nonlocality \cite{ida_2015,ida_2022}, the radial propagation of the fluctuations, or turbulence spreading \cite{hahm_2005} has been extensively studied in the past several decades.
Turbulence spreading can transfer free turbulent energy from strongly driven to weakly driven regions, and redistribute the turbulence intensity field.
For the current devices with moderate size, the derivation of transport scaling from gyro-Bohm scaling has been widely observed in the experiments \cite{petty_1995} and numerical simulations \cite{linz_2004,gorler_2011}.
Turbulence spreading is believed to contribute to that through the mesoscale dynamics \cite{hahm_2018,garbet_1994,gurcan_2005,yagi_2006}, such as linear toroidal coupling and nonlinear coupling between zonal flow and turbulence, which are lost in a local or quasilocal model \cite{callen_1992}.
In this sense, theory with mesoscale dynamics should be established to include the turbulence intensity or transport flux, profile modification, shear flow and other nonlocal effects.

An important application of turbulence spreading theory is the dynamics of edge-core interaction and coupling \cite{hahm_2005,singh_2020}, which are traditionally treated as independent.
For the L-mode plasmas, the turbulence intensity usually increases radially outward since the smaller scale length of density and temperature profiles at the edge and micro-instabilities are easier to excite.
The strong resistive turbulence in the edge or invasion of turbulence from the scrape-off layer could be also treated as the influx in turbulence spreading.
Therefore, the inward turbulence spreading is common in this situation, though the definition of edge-core boundary may be ambiguous.
For the H-mode plasmas, the quenching of turbulence in the core region which originated at the edge, during L-H transition can be found and the properties of turbulence spreading are thought to have a significant influence on the pedestal height and width.
On the other hand, during the H-L back transition, the fluctuations of turbulence can reach a high level accompanied by the disappearance of the edge transport barrier and the collapse of the edge flow shear.
Similarly, transport barrier also can be produced in the core region, which is referred as internal transport barrier(ITB).
Magnetic shear and flow shear are reported to be important for the ITB formation, and the properties of turbulence spreading in the weak magnetic shear region \cite{yi_2014} and through the strong flow shear region\cite{wangwx_2006,wangwx_2007} have been reported.
Moreover, the attention are also gained of the turbulence spreading across the magnetic island \cite{poli_2009,poli_2010} and stochastic magnetic layer \cite{kobayashi_2022}.

Since the multi-scale nature of plasma, the electric and magnetic field are also effected by other macro-scale structures, such as MHD activities.
As one of the dangerous MHD instabilities, edge localized modes (ELMs) \cite{lang_2013} can cause a partial collapse of the pedestal profiles in H-mode, thus induced the significant heat and particle fluxes.
Resonant magnetic perturbations (RMPs) have been shown to mitigate or suppress ELMs \cite{evans_2005,mckee_2013}, and it is reported that the radial electric field shear could be reduced at the pedestal top and turbulence spreading are enhanced \cite{holod_2017,taimourzadeh_2019}.
Furthermore, RMP is also used to modulate the tearing mode (TM) \cite{hender_1992,huqm_2012} and neoclassical tearing mode (NTM) \cite{haye_2006,tangwk_2020}, another dangerous MHD instability, and prevent the disruption.
Both the modification of radial electric field \cite{nishimura_2010} and magnetic island induced by the locked TM could affect the properties of turbulence and it has been studied with various electrostatic turbulence simulations.  

In this work, the effect of resonance magnetic perturbation on the inward turbulence spreading during the pedestal collapse are investigated based on the global electromagnetic Landau-fluid simulation \cite{ren_2022,lwei_2023}.
Different from the electrostatic simulations with the embedded magnetic island or radial electric field, the electromagnetic perturbation are calculated consistently.
The remainder of this paper is organized as follows. 
In Sec.~\ref{sec:model}, the global electromagnetic Landau-fluid model used in this work is briefly introduced. 
The characterizations of the inward propagation of the electromagnetic turbulence front is analyzed in Sec.~\ref{sec:spreading}. 
The effect of radial magnetic perturbation on the inward turbulence spreading is analyzed in Sec.~\ref{sec:blocking}.
Finally, a short summary and discussion are given in Sec.~\ref{sec:summary}.

\section{model} \label{sec:model}

Landau-fluid model is employed in this work, which consists of evolution equations for the perturbed density $\tilde{n}_e$, vorticity $\nabla_\perp^2\tilde{\phi}$, parallel ion velocity $\tilde{v}_{\parallel i}$, parallel magnetic vector potential $\tilde{A}_\parallel$, and ion temperature $\tilde{T}_i$,
\begin{equation}
\begin{aligned}
\frac{d}{dt} \tilde{n}_e
=
&- n_{eq}\nabla_\parallel \tilde{v}_{\parallel,i}
+ \frac{n_{eq}}{T_{i,eq}}i\omega_{\ast i}\tilde{\phi}
- \frac{n_{eq}}{T_{i,eq}}i\omega_{Di}\tilde{\phi}	\\
&+ \frac{1}{T_{i,eq}}i\omega_{Di}p_e
+ \nabla_\parallel j_\parallel
+ D_n\nabla_\perp^2 \tilde{n}_e,
\end{aligned}
\label{eq:ne}
\end{equation}
\begin{equation}
\begin{aligned}
\frac{d}{dt} \nabla_\perp^2\tilde{\phi}
=
&- i\omega_{\ast p_i}\nabla_\perp^2\tilde{\phi}
+ \frac{1}{p_{i,eq}}i\omega_{Di}(p_i+p_e)	\\
&+ \frac{1}{n_{eq}}\nabla_\parallel j_\parallel
+ D_U\nabla_\perp^4\phi,
\end{aligned}
\end{equation}
\begin{equation}
\frac{d}{dt} \tilde{v}_{\parallel,i}
=
- \frac{1}{n_{eq}}\nabla_\parallel\left( p_i+p_e \right)
+ D_v\nabla_\perp^2 \tilde{v}_{\parallel,i},
\end{equation}
\begin{equation}
\beta \frac{\partial}{\partial t}\tilde{A}_\parallel
=
- \nabla_\parallel\tilde{\phi}
+ \frac{1}{n_{eq}}\nabla_\parallel p_e
+ \sqrt{\frac{\pi}{2}\tau\frac{m_e}{m_i}}|\nabla_{\parallel}|\tilde{v}_{\parallel e}
-\eta \tilde{j}_\parallel,
\end{equation}
\begin{equation}
\begin{aligned}
\frac{d}{dt}\tilde{T}_i
=
&i\omega_{\ast T_i} \tilde{\phi}
- (\Gamma-1)T_{i,eq}\nabla_\parallel \tilde{v}_{\parallel,i}
- (\Gamma-1)i\omega_{Di}\tilde{\phi}	\\
&- (\Gamma-1)\frac{T_{i,eq}}{n_{eq}}i\omega_{Di} \tilde{n}_i
-(2\Gamma-1)i\omega_{Di}T_i	\\
&- (\Gamma-1)\sqrt{\frac{8T_{i,eq}}{\pi}}|\nabla_\parallel| \tilde{T}_i
+ D_T\nabla_\perp^2\tilde{T}_i.
\label{eq:ti}
\end{aligned}
\end{equation}
where,
\begin{equation}
	d_t\tilde{f}=\partial_t{\tilde{f}}+[\tilde{\phi},\tilde{f}]
	\nonumber
\end{equation}
\begin{equation}
	\nabla_\parallel\tilde{f}=\epsilon(\partial_\theta/q+\partial_\zeta)\tilde{f}
	-\beta[A_{\parallel,\mathrm{RMP}},\tilde{f}]
	-\beta[\tilde{A}_\parallel,\tilde{f}]
	\nonumber
\end{equation}
\begin{equation}
	i\omega_{\ast i}\tilde{f}
	= \frac{a}{r}\frac{T_{i,eq}}{n_{eq}}\frac{\partial n_{eq}}{\partial r}\frac{\partial\tilde{f}}{\partial\theta},\quad
	i\omega_{\ast T_i}\tilde{f}
	= \frac{a}{r}\frac{\partial T_{i,eq}}{\partial{r}}\frac{\partial\tilde{f}}{\partial\theta},\quad
	\nonumber
\end{equation}
\begin{equation}
	i\omega_{Di} \tilde{f} = -2\epsilon T_{i,eq}[r\cos\theta,\tilde{f}]
	\nonumber
\end{equation}
Here, the variables with subscript $eq$ means the time independent equilibrium variables.
Pressure $p_i = p_{i,eq}a + n_{eq} \tilde{T}_i + T_{i,eq} \tilde{n}_i$ and $p_e=p_{e,eq}a + T_{e,eq} \tilde{n}_e$ are the ion and electron pressure.
$\epsilon=a/R_0$ is the the inverse aspect ratio, $a$ and $R_0$ the normalized minor and major radius respectively.
$\tau=T_{e,eq}/T_{i,eq}$ is the ratio of electron and ion equilibrium temperatures.
Parallel current $\tilde{j}_\parallel = -\nabla_\perp^2 \tilde{A}_\parallel$, $j_{\parallel 0}=\nabla\times\bm{B}_0/\beta$ and $j_\parallel=j_{\parallel 0}a+\tilde{j}_\parallel+j_{\parallel,\mathrm{RMP}}$.
The circular tokamak geometry $(r,\theta,\zeta)$ is employed in the numerical simulation, where $r$ is the radius of magnetic surface, $\theta$ and $\zeta$ are poloidal and toroidal angles, respectively.
Poisson bracket $[\tilde{f},\tilde{g}]=(\partial_r{\tilde{f}}\partial_\theta{\tilde{g}}-\partial_r{\tilde{g}}\partial_\theta{\tilde{f}})/r$.
The variables in the equations are normalized as follows
\begin{equation}
\begin{aligned}
\frac{a}{\rho_i}
\left(
\frac{\tilde{n}_e}{n_{c}},
\frac{e\tilde{\phi}}{T_{c}},
\frac{\tilde{v}_{\parallel i}}{v_{ti}},
\frac{\tilde{A}_\parallel}{\beta{B_c}\rho_i},
\frac{\tilde{T}_i}{T_{c}}
\right) &\rightarrow
\left( \tilde{n}_e,\tilde{\phi},\tilde{v}_{\parallel i},\tilde{A}_\parallel,\tilde{T}_i
\right),	 \nonumber\\
\left( \frac{r}{\rho_i},
\rho_i\nabla_\perp,
a\nabla_\parallel,
\frac{v_{ti}}{a}t \right)	&\rightarrow
\left( r,\nabla_\perp,\nabla_\parallel,t \right).	\nonumber
\end{aligned}
\end{equation}
where $n_c$, $T_c$, $B_c$ are values at axis. 
$v_{ti}=\sqrt{T_c/m_i}$ and $\rho_i=m_iv_{ti}/eB_c$.
Parameter $\beta$ is defined as a half of the ratio of the thermal ion pressure to magnetic pressure in the center, $\beta = n_cT_c/(B_c^2/\mu_0)$.

\begin{figure}[H]
	\centering
	\includegraphics[width=0.4\textwidth]{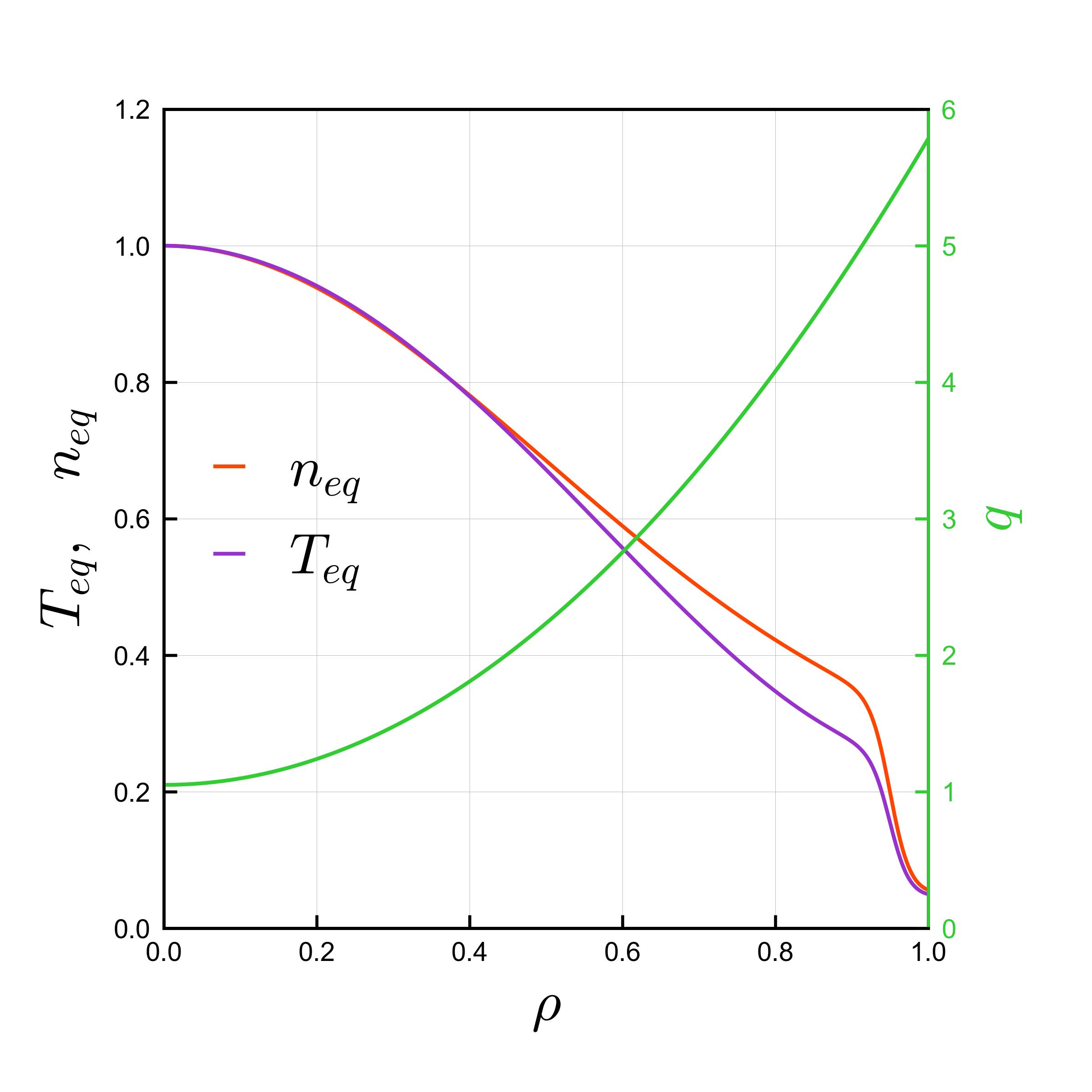}
	\caption{\label{fig:profile} The equilibrium profiles of safety factor $q$, density $n_{eq}$, temperature $T_{i,eq}$. }
\end{figure}

The Eqs.~(\ref{eq:ne})-(\ref{eq:ti}) is numerically solved as an initial value problem. 
Finite difference method is used in $r$ direction and the Fourier expansion in poloidal and toroidal directions. 
The inner boundary condition is given by
$\partial_r\tilde{f}_{0,0}|_{\rho=0}=0$ for $m=0,n=0$ mode and $\tilde{f}_{m,n}=0$ for $m\neq 0$ or $n\neq 0$.
Here $\rho=r/a$.
The zero Dirichlet condition is imposed for all perturbations on the outer boundary.
The equilibrium profiles of safety factor, density and ion temperature profiles are shown in Fig.~\ref{fig:profile}, where $n_{eq}$ and $T_{eq}$ are analytical expression combining the quadratic or exponential function at the core region and the hyperbolic tangent function at the edge, and $q$ is a simple quadratic function \cite{lwei_2023}.
Steep gradient exist in the edge density the temperature profiles, and the maximum scale lengths $R_0/L_T$ and $R_0/L_n$ are around 130 at edge. 
With steep gradient in the edge, turbulence spreading into the quiescent core region could be observed, and the effect of RMP field on the process of spreading is the main focus of this study. 
RMP filed is approximated by $A_{\parallel,\mathrm{RMP}}=A_{M,N}(r)=\psi_{a,M/N} A_{\parallel,eq}(r=a)\rho^M$, added to the equilibrium magnetic filed, where $\psi_a$ is the amplitude of filed  and $M$, $N$ the poloidal and toroidal mode number of RMP field respectively.

\section{characteristics of turbulence spreading} \label{sec:spreading}

\begin{figure}[h]
	\includegraphics[width=0.5\textwidth]{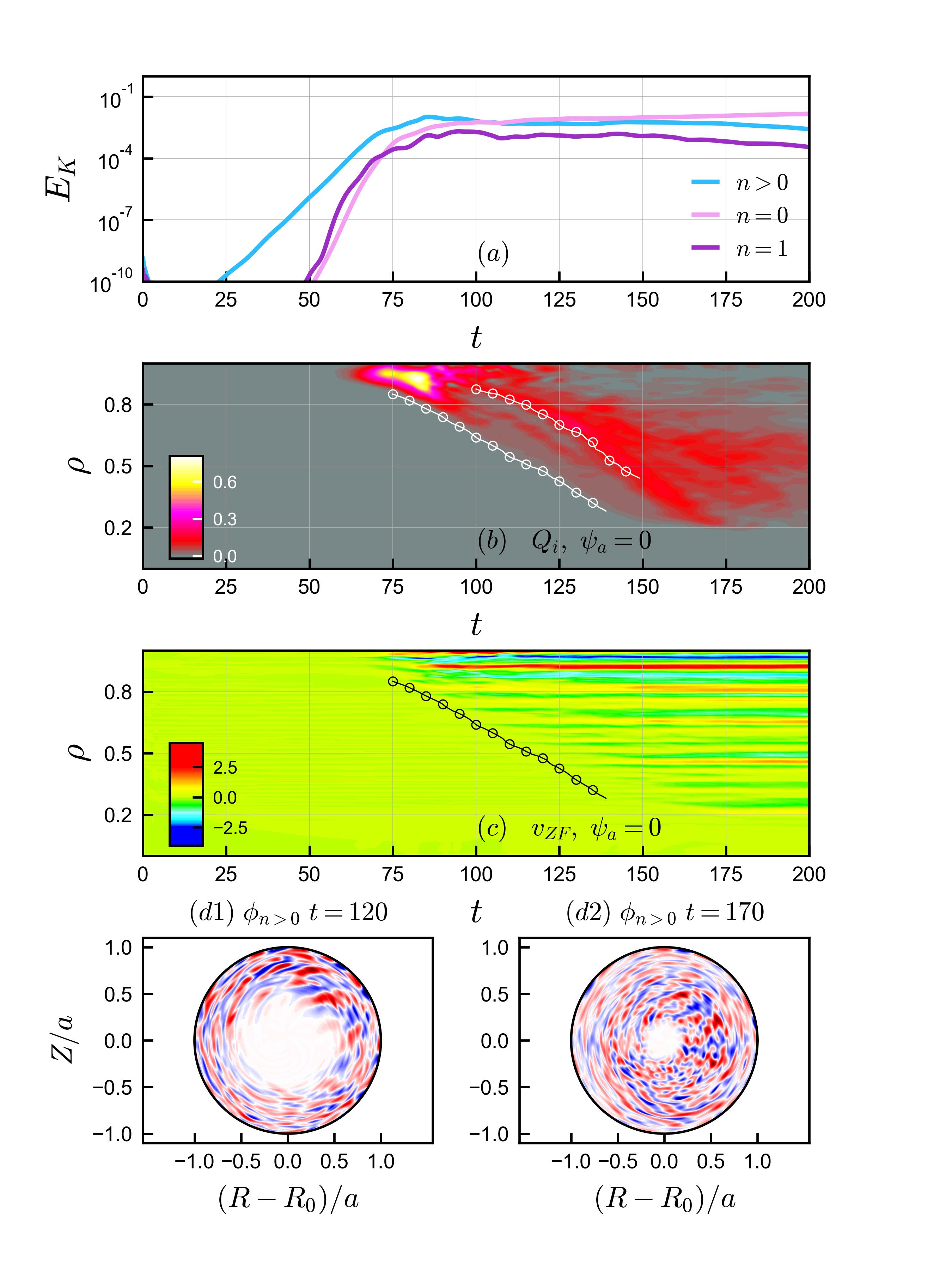}
	\caption{\label{fig:spreading} Time evolution of ion heat flux profile $(a)$ and zonal flow $(b)$. Electrostatic potential structures of turbulence at $t=120$ $(c1)$ and $t=170$ $(c2)$. }
\end{figure}

In the absence of RMP field, the dominant instabilities are the drift wave type instabilities driven by the steep density or temperature profiles, growing up firstly at the edge region and saturating before the subdominant instabilities growing up at the inner region.
The detailed analysis of the instabilities is provided in Fig.~2 and Fig.~3 in Ref.~\cite{lwei_2023}, which shows the dominant electron drift wave type instabilities at the edge, as well as the subdominant ion temperature gradient instabilities and kinetic ballooning mode at the core.
Fig.~\ref{fig:spreading}$(a)$ and $(b)$ give the temporal evolution of the kinematic energies of the perturbations with the kinematic energies $E_K = \int |\nabla_\perp\phi|^2 \rho d\rho/2$.
First, the fluctuations at edge grow up, zonal flows are excited and induce the saturation of turbulence around $t\sim70$.
After the saturation, the fluctuations start to spread into the core region as inward flux, as shown in Fig.~\ref{fig:spreading}$(b)$.
The ballistic propagation of the flux front are indicated by the white curve in Fig.~\ref{fig:spreading}$(b)$, which holds a nearly constant velocity, $u_r\sim 0.72 (\rho_i/a)v_{ti}$.
The linear coupling of poloidal harmonics and nonlinear coupling can contribute to the ballistic propagation of the flux front, as the theory given by Ref.~\cite{garbet_1994} and Ref.~\cite{hahm_2005}.

\begin{figure}[h]
	\includegraphics[width=0.5\textwidth]{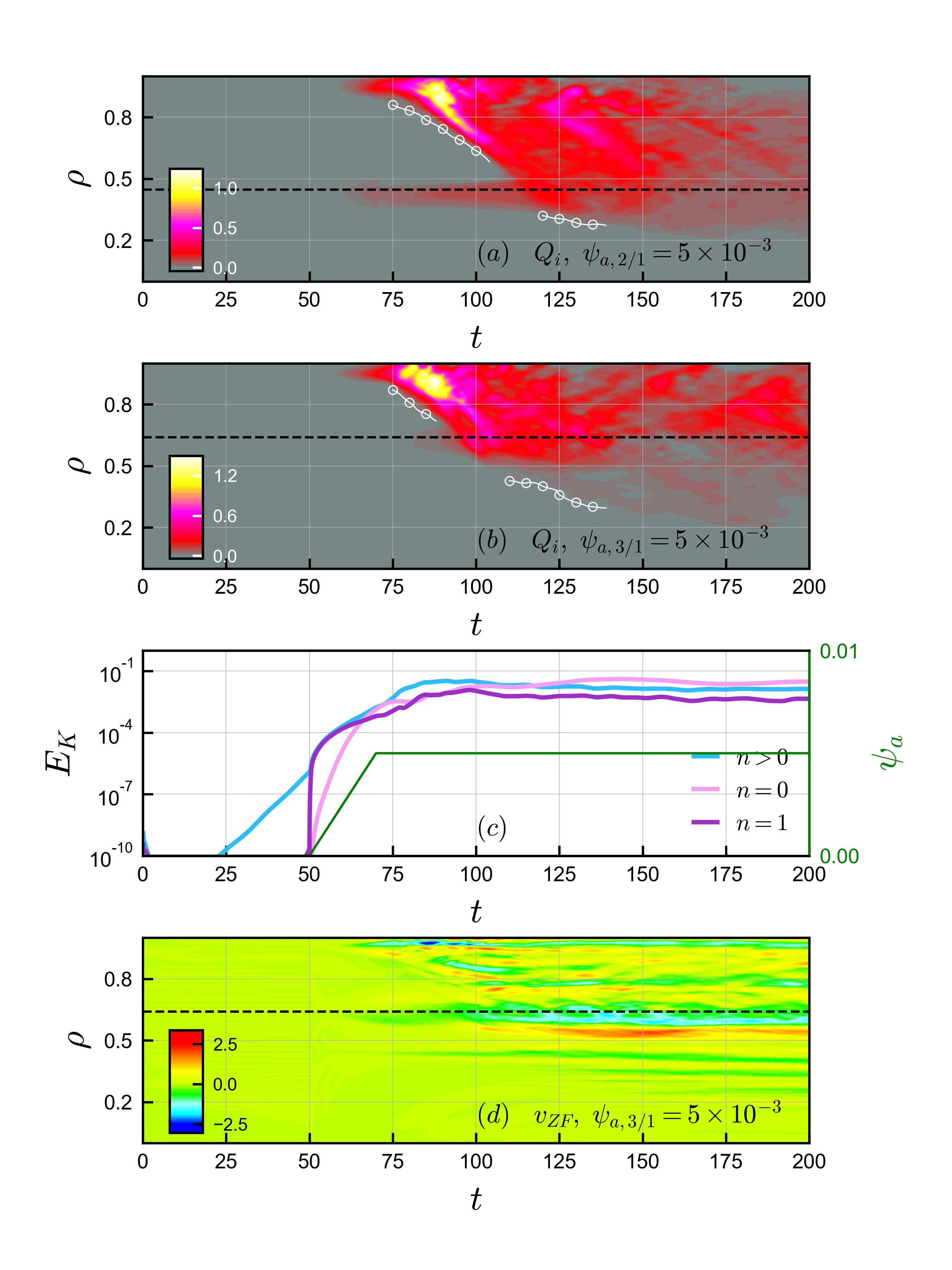}
	\caption{\label{fig:spreading_rmp} Time evolution of ion heat flux profile with $2/1$ RMP field and $3/1$ RMP field $(b)$. Time evolution of zonal flow $(c)$. }
\end{figure}

In Fig.~\ref{fig:spreading}$(b)$, the locations of $1\%$ or $2\%$ of the maximum of heat flux are indicated by the white curves.
It can be found that the inward spreading speeds of the turbulence are almost the same for the two cases with tracing the $1\%$ and $2\%$ of the maximum of heat flux profiles. 
The remarkable difference is that the spreading velocity gradually increases as the fluctuations propagates in the core region for the case with tracing the $2\%$ of the maximum of heat flux.
This phenomenon may be caused by the features of the instabilities at the core region, since for the core fluctuations, the streamer structures are more prevalent, which could enhance the turbulence intensity, comparing to the edge instabilities. 
Radial elongated vortex structures can be observed in the process of spreading, as shown in Fig.~\ref{fig:spreading}$(d1)$.
Since no heat source are considered in the present study, profile relaxation occurs after the turbulence saturation at the edge region, and the fluctuations are mainly concentrated in the core region after the profile relaxation, as shown in Fig.~\ref{fig:spreading}$(d2)$. 
Thus, one could see the larger saturation level when the influx arrives the core region.
The time evolution of zonal flow is given in Fig.~\ref{fig:spreading}$(c)$.
The zonal flow structures also have the same inward spreading speed as the turbulence fluctuations, as indicated by the black curve.
$\bm{E}\times\bm{B}$ staircase can be observed, and in the core region, zonal flow is weaker, which also contributes to the high saturation level of turbulence.

\section{blocking of spreading} \label{sec:blocking}

\begin{figure}[h]
	\centering
	\includegraphics[width=0.5\textwidth]{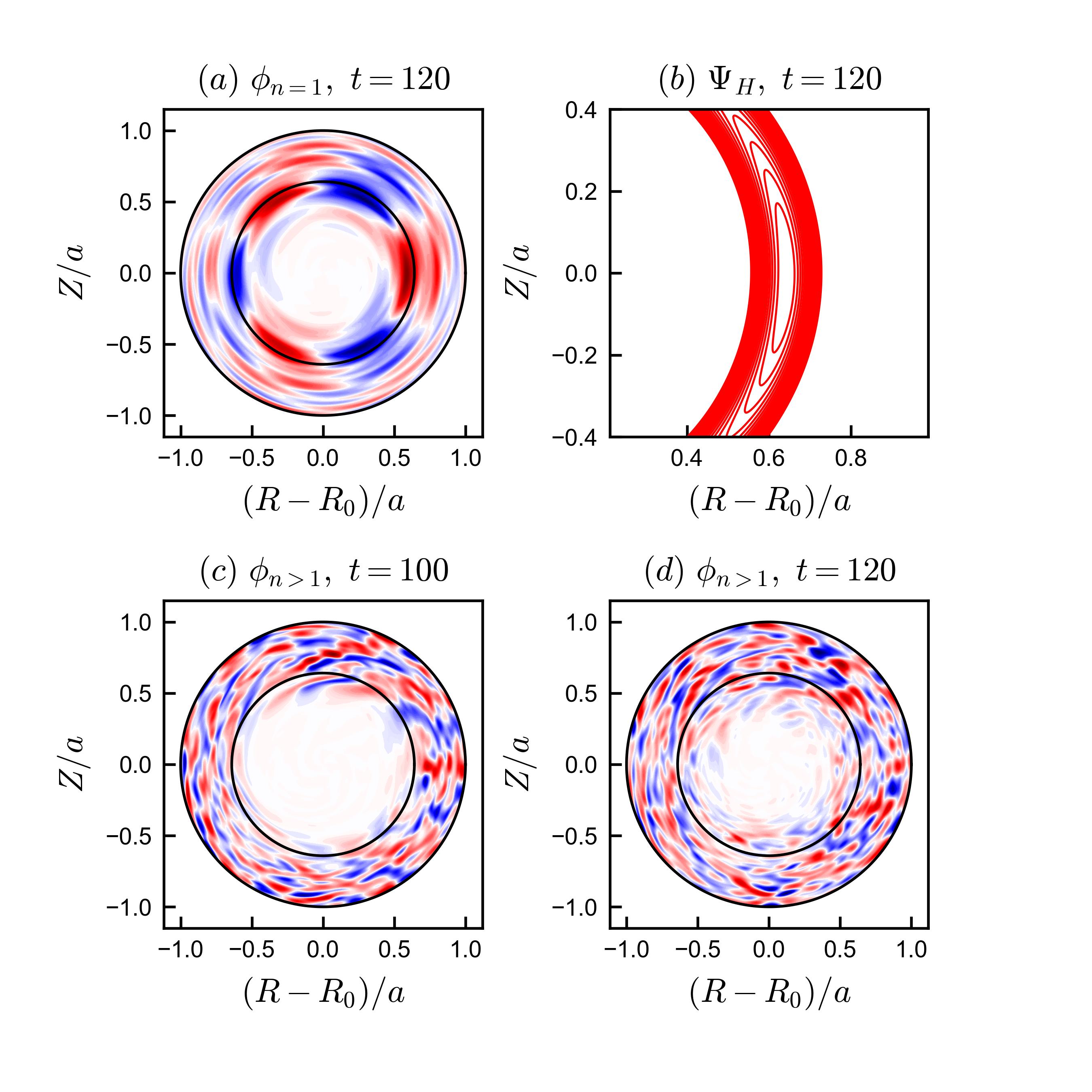}
	\caption{\label{fig:contour} Electrostatic potential structures with $3/1$ RMP field of $\phi_{n=1}$ at $t=120$ $(a)$, $\phi_{n>1}$ at $t=100$ $(c)$ and $\phi_{n>1}$ at $t=120$ $(d)$. The helical function with $3/1$ perturbations is given in $(b)$. }
\end{figure}

\begin{figure}[h]
	\centering
	\includegraphics[width=0.5\textwidth]{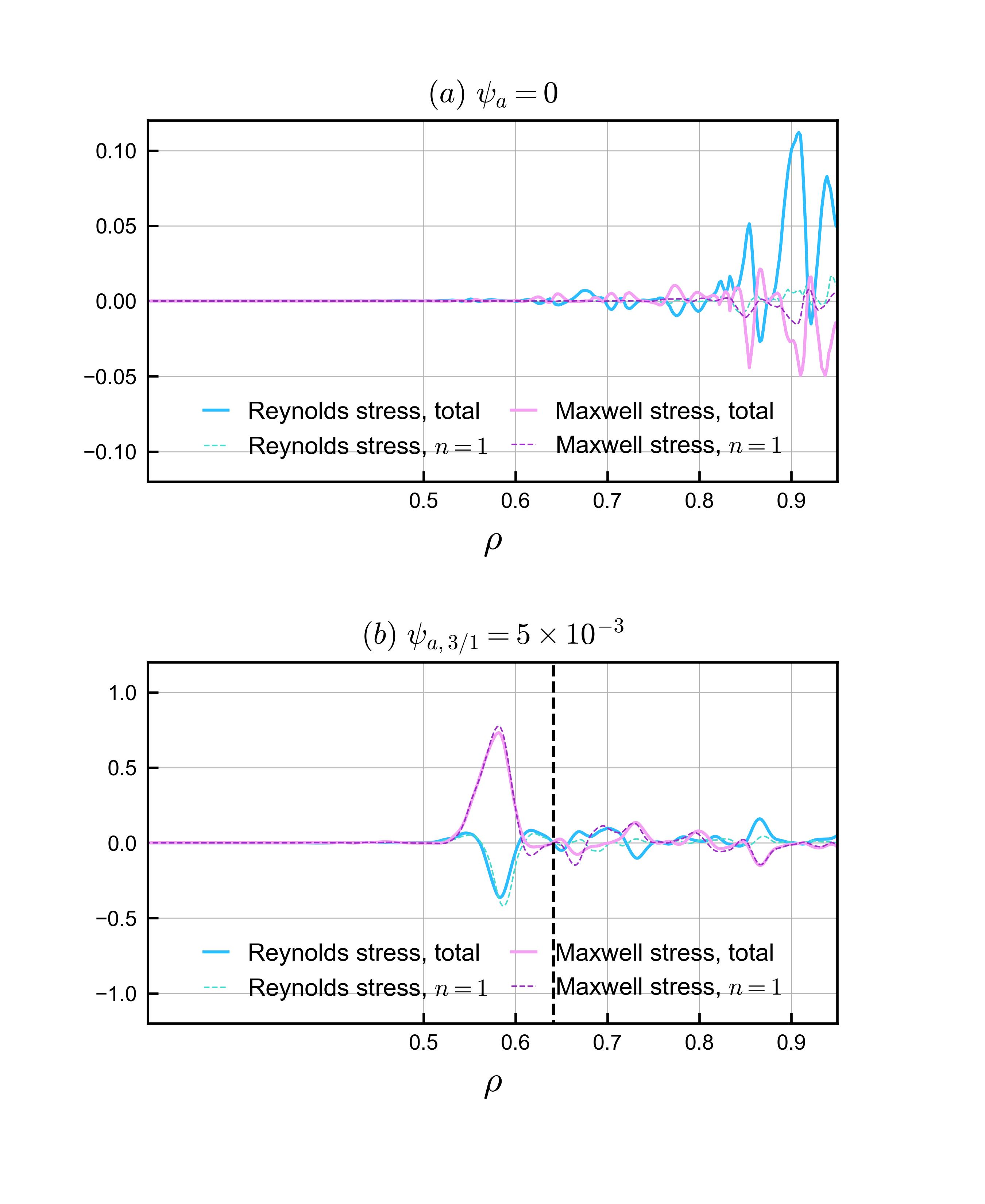}
	\caption{\label{fig:zf_drive} Reynolds stress and Maxwell stress distribution averaged during $t\in[100,150]$ with no RMP field $(a)$ and with $3/1$ RMP field $(b)$.}
\end{figure}

When the appropriate $\psi_a$ is set, the RMP field could possess an impact on the turbulence spreading between the edge and core regions.
As shown in Fig.~\ref{fig:spreading_rmp}$(a)$ and $(b)$, the turbulence stop spreading as $2/1$ or $3/2$ RMP field is employed.
The RMP field has little effect on the initial propagation velocity, and the fluctuations propagate at the same rate as in the absence of RMP, as indicated by flux front locations with white curves in Fig.~\ref{fig:spreading_rmp}.
When the turbulence arrives the corresponding resonance surface with RMP field, which is indicated by the black dash line in Fig.~\ref{fig:spreading_rmp}, the properties of the spreading change.
First, as indicated by the front locations when $t\in(110,140)$, the velocity of turbulence spreading obviously declines.
Second, the amplitude of heat flux at the inner region of resonance surface also decrease comparing to the no RMP case, indicating that the saturation level of turbulence in the core region decreases due to the effect of the RMP field.
The variation of the electric and magnetic fields around the resonance surfaces is the cause of the change of the turbulence propagation characteristics.

As one typical plasma response to the RMP field, forced magnetic reconnection could occur around the resonance surface as the resonant response.
The long wave length fluctuations thus grow, as shown in Fig.~\ref{fig:spreading_rmp}$(c)$ and Fig.~\ref{fig:contour}$(a)$, the formation of magnetic island and modification of local radial electric field both make the possible impact on the turbulence.
An interesting observation is that strong shear flow is induced around the resonance surface, as shown in Fig.~\ref{fig:spreading_rmp}$(d)$ for the $3/1$ RMP case.
Comparing to the case in the absence of RMP, the shear flow around the $q=3$ surface is excited before the turbulence front arrives and the amplitude of that is higher than the shear flow at edge.
The helical function with $3/1$ perturbations is given in Fig.~\ref{fig:contour}$(b)$ and the obvious magnetic island can be seen with a width of several $\rho_i$.
Actually, with this magnetic island, the shear flow is mainly excited at the location of the inner separatrix of the island, similar to the Ref.~\cite{nishimura_2010}. 
This strong shear flow reduces the scale length of the turbulent eddies, as shown in Fig.~\ref{fig:contour}$(c)$.
As one can see, around the $q=3$ surface, the radially elongated eigenmode structures in the absence of RMP are broken up, and the decline of radial length of turbulence fluctuation could decrease the transport, thus blocking the turbulence spreading to the inner region.
As a result, the fluctuations are mainly concentrated in the outer region of $q=3$ surface and the saturation level of fluctuations decrease a lot in the inner region compared to the case with no RMP filed, as shown in Fig.~\ref{fig:contour}$(d)$.
Correspondingly, the amplitude of $\phi_{0,0}$ is also small in the inner side of $q=3$ surface.

To validate the driven mechanism of zonal flow with RMP field, the drive stress analysis are carried out, as shown in Fig.~\ref{fig:zf_drive}.
In the absence of RMP field, since the steep gradient in the edge, zonal flow is primarily driven by the Reynolds stress at the edge region when $t\in[100,130]$.
Dominated fluctuations hold $n>1$ mode number and broad spectra contribute to the zonal flow excitation.
Maxwell stress contributes to the excitation of zonal flow as sink term.
With RMP field, after the long wave length fluctuations grow up, zonal flow is strongly excited at the separatrix of the magnetic island.
Correspondingly, the Reynolds stress and Maxwell stress both peak and $n=1$ mode contribution dominates the component of them.
Interestingly, zonal flow at the separatrix of the magnetic island is mainly driven by the Maxwell stress, indicating strong magnetic fluctuations induced by the RMP, and the energy of long wave length fluctuations are transferred to other turbulence fluctuations by the Reynolds stress. 
Moreover, due to the coupling of poloidal harmonics for $n=1$ mode, the excitation of zonal flow in the other regions is also dominated by $n=1$ component, which destroys the staircase structures of zonal flow in the absence of RMP, as shown in Fig.~\ref{fig:spreading_rmp}$(d)$.

\section{conclusion} \label{sec:summary}

In summary, we have presented global electromagnetic Landau-fluid simulation results of the effect of RMP on edge to core turbulence spreading.
In the absence of RMP field, ballistic propagation of turbulence are observed, which holds the nearly constant velocity.
After the profile relaxation in the edge region, saturation turbulence are mainly concentrated in the core region.
RMP field could affect the process of turbulence spreading though the excitation of the long wave length fluctuations due to the forced magnetic reconnection.
When RMP field with appropriate amplitude are employed in the simulation, obvious magnetic island along with the strong shear flow at the island separatrix can be observed.
The strong shear flow is identified as induced by Maxwell stress dominated by $n=1$ mode.
Moreover, the turbulence spreading could be blocked by this strong zonal flow shear.
The radial elongated vortex structures are broken, and the transport level are reduced.
With this blocking effect, the saturation level in the core region also declines a lot. 
These qualitative features appear to be robust consequences of our work and have potentially consequences for the multi-scale interaction among micro turbulence, zonal flow and long wave length MHD mode.

\begin{acknowledgments}

This work is supported by the National MCF Energy R\&D Program of China (Grant Nos. 2019YFE03090300 and 2017YFE0301100), National Natural Science Foundation of China (Grant Nos. 11925501, 12275071, 12075048 and U1967206), the Fundamental Research Funds for the Central Universities (Grant No. DUT21LK28). The calculations of this work is supported by Supercomputing Center of Dalian University of Technology. The calculation is also carried out on Tianhe-3 prototype at Chinese National Supercomputer Center in Tianjin.

\end{acknowledgments}

\bibliography{ref}

\end{document}